\documentclass[conference,a4paper]{IEEEtran}
\IEEEoverridecommandlockouts

\usepackage[hidelinks]{hyperref}
\usepackage[cmex10]{amsmath}
\usepackage{amssymb,amsfonts}
\usepackage{dblfloatfix}

\usepackage[ruled,vlined]{algorithm2e}
\usepackage{graphicx}
\graphicspath{{Figures/PDF/}{Figures/PNG/}}

\usepackage{booktabs}
\usepackage{siunitx}
\usepackage[numbers,compress]{natbib}
\usepackage{texnames}
\usepackage{bm,bbm}
\usepackage{orcidlink}
\usepackage{subcaption}
\usepackage{multirow}

\newcommand{\Fig}{Fig.\@\xspace}

\begin{document}

\title{\uppercase{Digital Elevation Model Estimation from RGB Satellite Imagery using Generative Deep Learning}
\thanks{This research was funded by Google Research Scholar Program and Ford Foundation. *Corresponding author: risqi.saputra@monash.edu.}
}

\author{
    \IEEEauthorblockN{
        Alif Ilham Madani\orcidlink{0009-0009-7894-745X},  
        Riska A. Kuswati\orcidlink{0009-0006-8332-5514},
        Alex M. Lechner\orcidlink{0000-0003-2050-9480}, 
        Muhamad Risqi U. Saputra*\orcidlink{0000-0001-6067-630X}
    }
    \IEEEauthorblockA{
        \textit{Monash University Indonesia}, 15345 Tangerang Selatan, Indonesia\\
    }
}

\maketitle
\begin{abstract}
Digital Elevation Models (DEMs) are vital datasets for geospatial applications such as hydrological modeling and environmental monitoring. However, conventional methods to generate DEM, such as using LiDAR and photogrammetry, require specific types of data that are often inaccessible in resource-constrained settings. To alleviate this problem, this study proposes an approach to generate DEM from freely available RGB satellite imagery using generative deep learning, particularly based on a conditional Generative Adversarial Network (GAN). We first developed a global dataset consisting of 12K RGB-DEM pairs using Landsat satellite imagery and NASA's SRTM digital elevation data, both from the year 2000. A unique preprocessing pipeline was implemented to select high-quality, cloud-free regions and aggregate normalized RGB composites from Landsat imagery. Additionally, the model was trained in a two-stage process, where it was first trained on the complete dataset and then fine-tuned on high-quality samples filtered by Structural Similarity Index Measure (SSIM) values to improve performance on challenging terrains. The results demonstrate promising performance in mountainous regions, achieving an overall mean root-mean-square error (RMSE) of 0.4671 and a mean SSIM score of 0.2065 (scale -1 to 1), while highlighting limitations in lowland and residential areas. This study underscores the importance of meticulous preprocessing and iterative refinement in generative modeling for DEM generation, offering a cost-effective and adaptive alternative to conventional methods while emphasizing the challenge of generalization across diverse terrains worldwide.
\end{abstract}



\section{Introduction}

Digital Elevation Models (DEMs) play a crucial role in geospatial analysis, enabling applications in hydrological modeling, environmental monitoring, and disaster management. Traditional methods for generating DEMs, such as those based on LiDAR \cite{aguilar2010modelling, pinton2021estimating} or stereo-photogrammetry \cite{st2008mapping}, require expensive equipment and extensive and specialized data collection (i.e. overlapping images), making them inaccessible in resource-limited settings. This has driven research toward leveraging publicly available satellite imagery to generate DEMs through cost-effective machine learning techniques.

One of the challenges in using RGB imagery to estimate DEM is the inherent difficulty in representing elevation data due to the lack of direct height information in visible-spectrum images. The conventional approach using stereo-photogrammetry requires two or more overlapping RGB images and ground control points to calibrate the images and accurately estimate elevation \cite{st2008mapping}. However, this information is often unavailable in publicly accessible satellite imagery, such as from the Landsat and the Sentinel Earth Observation Satellite. Previous studies, such as \cite{panagiotou2020generating}, have explored generating DEMs using remotely sensed RGB images and deep learning techniques. However, these models were trained and tested on images captured in a constrained region, namely Greece, which raises questions about their generalization performance and evaluation across diverse terrains globally.


This study presents an approach for generating DEMs from globally acquired RGB satellite imagery, specifically from Landsat satellites, using generative deep learning. We apply a conditional Generative Adversarial Network (GAN), specifically the \textit{pix2pix} architecture \cite{pix2pix}, for DEM generation. To enhance the DEM generation process, we propose a unique methodology. First, preprocessing is performed to ensure high-quality training data by selecting cloud-free locations and preprocessing to create cloud-free multi-date mosaics. The model then undergoes a two-stage training process: the model is initially trained on the entire dataset, followed by fine-tuning using samples filtered based on their Structural Similarity Index Measure (SSIM) \cite{hore2010image} values, refining the network’s ability to generate more accurate DEMs. This iterative process improves model performance in challenging terrains and demonstrates the utility of integrating quality-driven feedback loops. Both quantitative and qualitative evaluations are performed to assess the model’s performance across diverse regions, providing valuable insights and lessons for future research directions. 


\section{Data and Methodology}

\subsection{Data}
This study utilizes RGB and DEM images from publicly available satellite datasets to develop the model, all from the year 2000. The training data consist of 12K carefully selected paired RGB-DEM samples. The data sources and timeframes are described as follows:

\begin{itemize}
    \item \textbf{Cloud Fraction Data:} Monthly cloud fraction data were obtained from Terra/MODIS \cite{modis_cloud_fraction} for February 2000. This dataset was used to identify locations with no cloud coverage.
    \item \textbf{RGB Data:} Landsat 5 \cite{landsat_5} and Landsat 7 \cite{landsat_7} Level 2, Collection 2, Tier 1 images were accessed via Google Earth Engine for the time period from January 1, 2000, to December 31, 2000. These data are preprocessed and used as input images to the conditional GAN.
    \item \textbf{DEM Data:} The DEM data were sourced from NASA's Shuttle Radar Topography Mission (SRTM) Digital Elevation dataset \cite{srtm_dem}, with a spatial resolution of 30 meters, corresponding to the period from February 11, 2000, to February 22, 2000. These data serve as the ground truth DEM to train and test the model.
\end{itemize}

\begin{figure}[h!]
\centering
\includegraphics[width=0.95\columnwidth]{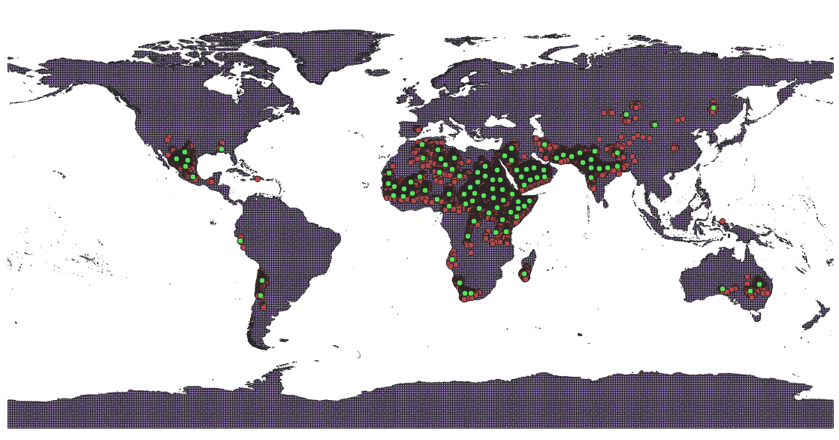}
\caption{Global map showing the locations of the data used in this study. Red dots represent the locations of individual image samples, while green dots indicate the centroids of the k-means clusters.}
\label{fig:data_world_map}
\end{figure}

\subsection{Preprocessing Steps}

Preprocessing was conducted to ensure high-quality data for training and validation. To identify cloud-free regions, cloud fraction data from Terra/MODIS was analyzed, and coordinates with 0\% cloud coverage were extracted. A 0.135-degree buffer was applied to each coordinate, defining regions as \texttt{(lat, lon) ± 0.135}. To further refine the selection, k-means clustering was employed with a batch size of 100 clusters, generating representative clusters for diverse spatial coverage as shown in \Fig~\ref{fig:data_world_map}.

For RGB data extraction, imagery from Landsat 5 and Landsat 7 was processed for the date range of January 1, 2000, to December 31, 2000. Images were filtered to include only those with less than 20\% cloud cover and were subsequently cleaned using a cloud and cloud-shadow mask. The resulting data were aggregated by calculating the median value of pixel intensities to create a multi-date, cloud-free mosaic. Landsat 5 data were prioritized, and in cases where no valid data were available, imagery from Landsat 7 was utilized as a fallback.

DEM data were extracted for the selected cloud-free regions using the NASA SRTM dataset. This dataset provides elevation data at a spatial resolution of 30 meters, ensuring accurate representation of terrain features. The DEM data were processed for the same geographic regions and timeframes as the RGB data to maintain consistency in the study. Additionally, the elevation values were converted to relative elevation to account for local topographic variations. 

To further refine the dataset, an additional filtering process was applied to remove low-quality samples. This involved two criteria based on pixel distribution in the RGB images. First, images where any spectral band had fewer than 20 unique pixel values were flagged, as such low variation indicates poor data quality. Second, images where more than 20\% of total pixels shared a single unique value were also flagged. These thresholds ensured the removal of images with limited spectral diversity or significant noise. A union of both flagged sets was created, and the corresponding RGB-DEM pairs were excluded from the dataset. After this filtering process, the dataset was reduced from the original 15,000 image pairs to 12,357 high-quality pairs, ready for model training.

Both RGB and DEM data were then divided into training (80\%), validation (10\%), and test sets (10\%) to support model development. To ensure compatibility with the training process, a two-step normalization was applied to both datasets. Given RGB $x \in {\rm I\!R}^{( w \times h \times 3)}$ and DEM $y \in {\rm I\!R}^{( w \times h \times 1)}$ images, both raw data were converted to the JPEG range [0–255] using the Stretch Min-Max method as described below:
\begin{equation}
P_{\text{normalized}} = 255 \cdot \frac{P - P_{\text{min}}}{P_{\text{max}} - P_{\text{min}}}, 
\end{equation}
where $P$ represents every pixel value from both $x$ and $y$. Note that $P_{\text{min}}$ and $P_{\text{max}}$ describe the lower and upper percentiles of the data, typically set to the 2nd and 98th percentiles, respectively. Values below $P_{\text{min}}$ were set to 0, and values above $P_{\text{max}}$ were set to 255.

In the second step, the normalized JPEG values were scaled to the range [-1, 1] to prepare the data for model training. This was achieved using the following equation:
\begin{equation}
P_{\text{scaled}} = \frac{P_{\text{normalized}}}{127.5} - 1,
\end{equation}
applied to both $x$ and $y$, converting them into normalized and scaled RGB $\hat{x}$ and DEM $\hat{y}$ images. This two-step process ensured that both the RGB and DEM datasets were uniformly scaled and ready for training, enabling the model to efficiently process and learn from the data.

\subsection{Model Training}

The methodology employed in this study focuses on image-to-image translation using a conditional GAN framework, specifically based on \textit{pix2pix} \cite{pix2pix}. In our case, the objective of the conditional GAN is to generate a DEM image from a random noise vector \(z\) conditioned on the input RGB image $\hat{x}$. Similarly to the standard GAN's training procedure, we trained two neural networks, the generator \(G(\hat{x}, z)\) and the discriminator \(D(\hat{x}, \hat{y})\), in a mini-max game to enable the generator to produce output that mimics the real data distribution. The generator \(G(\hat{x}, z)\) learns a mapping from \((\hat{x}, z)\) to $\hat{y}$, while the discriminator \(D(\hat{x}, \hat{y})\) evaluates whether a given pair \((\hat{x}, \hat{y})\) is real or fake. By iteratively refining the generator and discriminator, the generator learns to produce output (fake DEM) that closely resembles the target DEM, ultimately fooling the discriminator into classifying them as the real target DEM.

To train the conditional GAN, we leverage two objective functions, namely generator and discriminator losses. The generator loss $L_{\text{gen}}$ combines the GAN loss $L_{\text{GAN}}$, which ensures the generator outputs are classified as real by the discriminator, and the \(L_1\)-loss, which minimizes the pixel difference between the generated output and the target output. $L_{\text{gen}}$ is defined as follows:

\begin{align}
\begin{split}
& L_{\text{gen}} = L_{\text{GAN}} + \lambda L_1, \\
& L_{\text{GAN}} = \mathbb{E}_{\hat{x}, z} \left[ L_{\text{BCE}}(1, D(\hat{x}, G(\hat{x}, z))) \right], \\
& L_1 = \frac{1}{N} \sum_{i=1}^{N} \|G(\hat{x}, z)_i - \hat{y}_i\|_1
\end{split}
\end{align}
where \(L_{\text{BCE}}\) measures the binary cross-entropy loss between the discriminator’s prediction $D(\hat{x}, G(\hat{x}, z))$ and the label \(1\) (real DEM), \(N\) is the number of pixels, and \(\lambda\) represents a weighting factor used to balance the GAN loss and \(L_1\) loss. On the other hand, since the discriminator aims to correctly classify real pairs \((\hat{x}, \hat{y})\) as real and generated pairs \((\hat{x}, G(\hat{x}, z))\) as fake, the discriminator loss is then defined as
\begin{align}
\begin{split}
L_{\text{disc}} = \mathbb{E}_{\hat{x}, \hat{y}} \left[ L_{\text{BCE}}(1, D(\hat{x}, \hat{y})) \right] + \\ \mathbb{E}_{\hat{x}, z} \left[ L_{\text{BCE}}(0, D(\hat{x}, G(\hat{x}, z))) \right]
\end{split}
\end{align}
In this case, \(L_{\text{BCE}}(1, D(\hat{x}, \hat{y}))\) ensures that real pairs \((\hat{x}, \hat{y})\) are classified as real (label \(1\)), while \(L_{\text{BCE}}(0, D(\hat{x}, G(\hat{x}, z)))\) ensures that generated pairs \((\hat{x}, G(\hat{x}, z))\) are classified as fake (label \(0\)). By alternately training using Equations (3) and (4), the model is expected to converge, enabling the generator to produce outputs that closely resemble the target DEM.

During the training process, we introduced two unique stages. In the first stage, the model was trained on the complete dataset using a learning rate of \(2 \times 10^{-4}\) for 1000k epochs. The performance of the model was then evaluated using the Structural Similarity Index Measure (SSIM) with a scale of -1 to 1, which quantifies the similarity between the generated and ground truth DEMs. Based on the evaluation results, samples with SSIM scores below thresholds of 0.1, 0.2, and 0.3 were discarded for further fine-tuning.

In the second stage, the model was retrained by removing low-performing samples while keeping the test set unchanged to ensure consistent evaluation. This refinement process aimed to eliminate data points that could introduce noise during training, allowing the model to focus on higher-quality samples. A lower learning rate of \(1 \times 10^{-4}\) is used, and training is conducted for 500k epochs. By reducing the influence of noisy or low-quality data, the model demonstrated enhanced stability and improved generalization, particularly in challenging regions that previously exhibited suboptimal performance.

\begin{figure}[h!]
    \centering
    \begin{subfigure}{0.3\linewidth}
        \includegraphics[width=\linewidth]{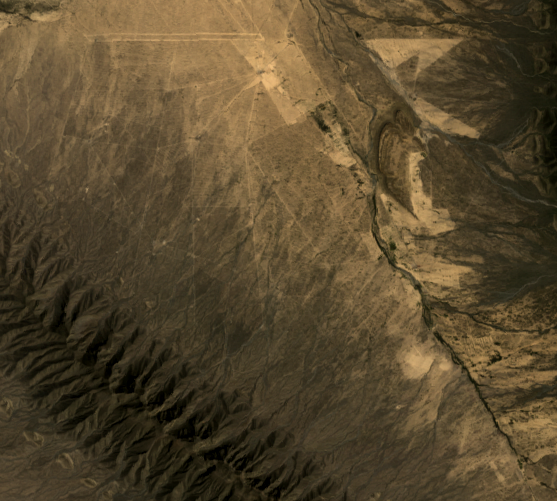}
        
    \end{subfigure}
    \hfill
    \begin{subfigure}{0.3\linewidth}
        \includegraphics[width=\linewidth]{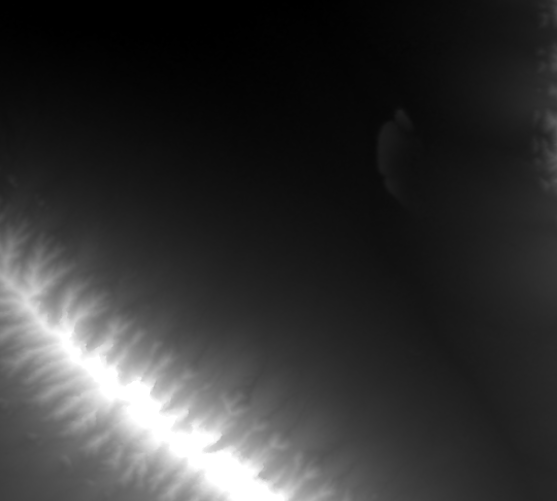}
       
    \end{subfigure}
    \hfill
    \begin{subfigure}{0.3\linewidth}
        \includegraphics[width=\linewidth]{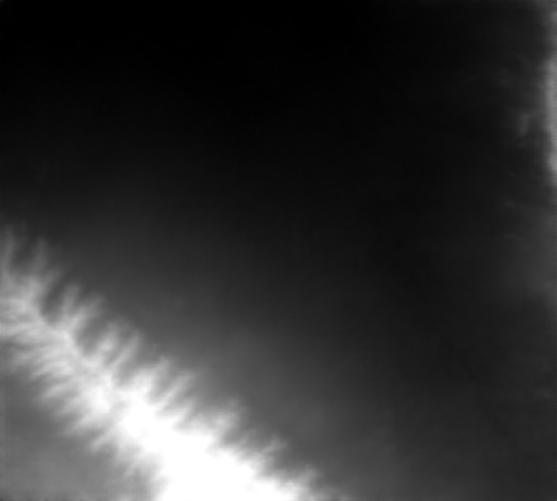}
        
    \end{subfigure}
    \begin{subfigure}{0.3\linewidth}
        \includegraphics[width=\linewidth]{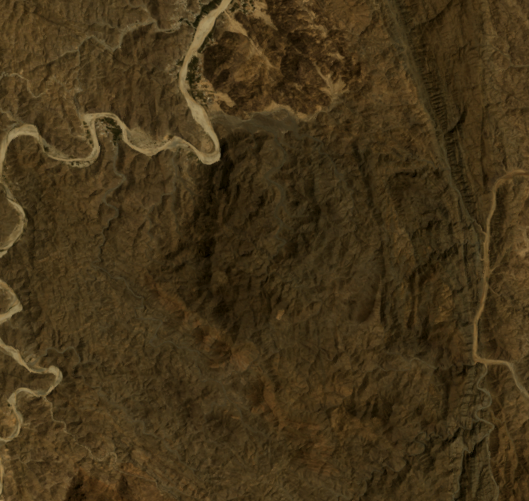}
    \end{subfigure}
    \hfill
    \begin{subfigure}{0.3\linewidth}
        \includegraphics[width=\linewidth]{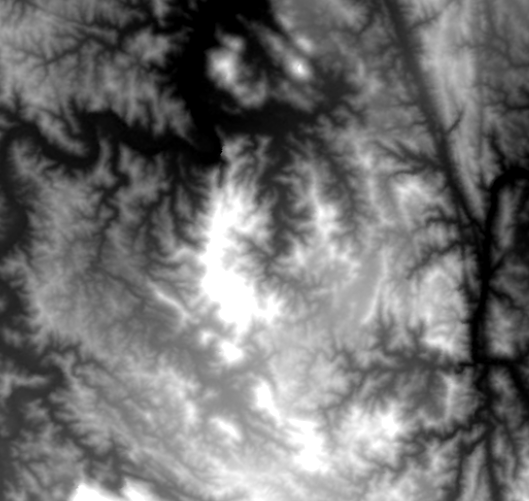}
    \end{subfigure}
    \hfill
    \begin{subfigure}{0.3\linewidth}
        \includegraphics[width=\linewidth]{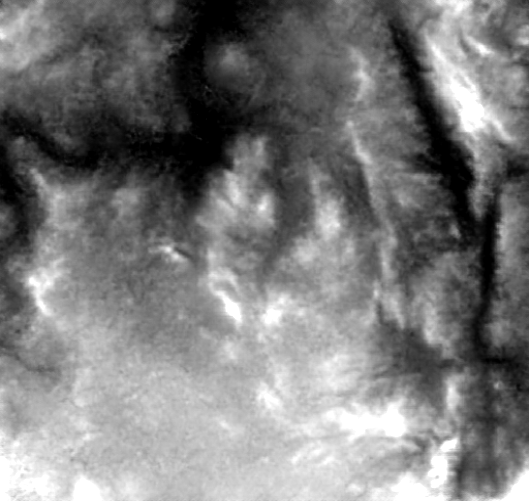}
    \end{subfigure}
    \begin{subfigure}{0.3\linewidth}
        \includegraphics[width=\linewidth]{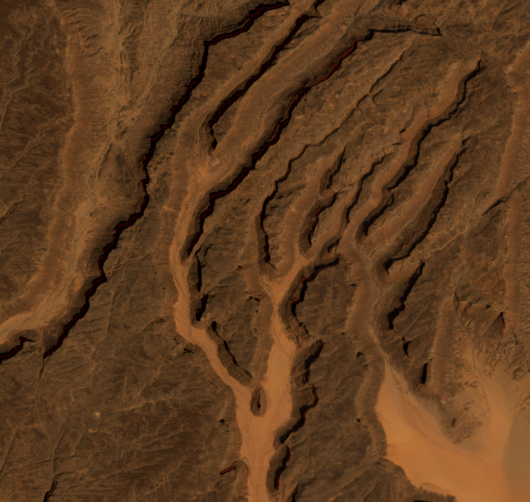}
        \caption{RGB Input}
    \end{subfigure}
    \hfill
    \begin{subfigure}{0.3\linewidth}
        \includegraphics[width=\linewidth]{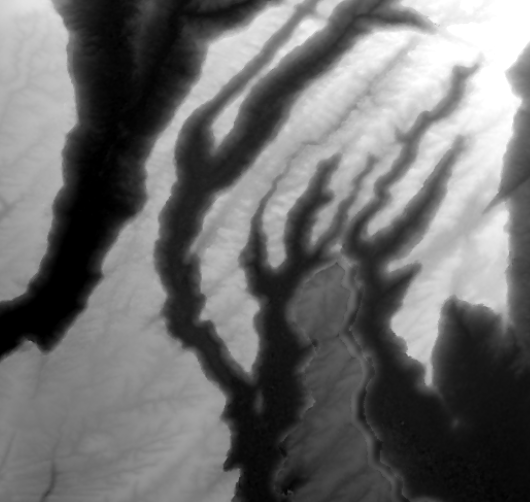}
        \caption{GT DEM}
    \end{subfigure}
    \hfill
    \begin{subfigure}{0.3\linewidth}
        \includegraphics[width=\linewidth]{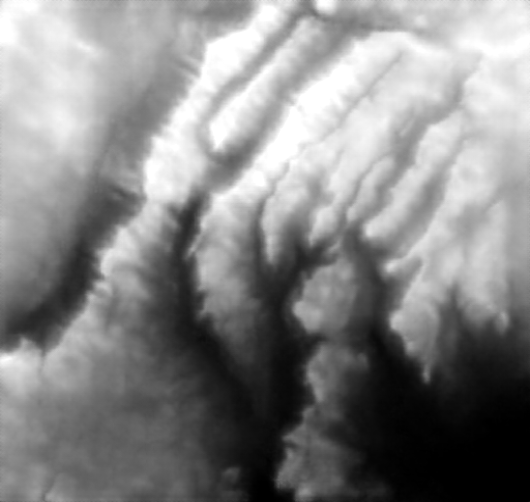}
        \caption{DEM Prediction}
    \end{subfigure}
    \caption{Sample images showing (a) RGB input, (b) ground truth (GT) DEM, and (c) predicted DEM on the test set.}
    \label{fig:sample_images}
\end{figure}

\section{Results and Discussion}

\subsection{Results}

The proposed method was evaluated using per-pixel RMSE and SSIM to assess the performance of the model in generating DEMs from RGB imagery. The RMSE was calculated on a pixel value scale ranging from -1 to 1. The initial 12K paired RGB-DEM samples served as the baseline dataset, on which the model achieved a mean SSIM of 0.1896 and a mean RMSE of 0.4876 (Table~\ref{tab:eval_stats}). After fine-tuning with filtered datasets, the performance improved. Filtering at SSIM $\geq$ 0.2 resulted in a mean SSIM of 0.2065 and a mean RMSE of 0.4671, demonstrating better reconstruction accuracy and structural similarity to the ground truth.

Filtering low-quality samples refined the dataset, enabling the model to focus on higher-quality data during training. This process reduced variability, as reflected in the consistent decrease in median RMSE values. However, residual noise in certain samples may account for smaller changes in mean RMSE. These findings highlight the balance between dataset filtering and retaining diversity for effective model training.

\begin{table}[h!]
\centering
\caption{Model Evaluation Results}
\label{tab:eval_stats}
\begin{tabular}{|l|cc|cc|}
\hline
\multirow{2}{*}{\textbf{Dataset}} & \multicolumn{2}{c|}{\textbf{SSIM}} & \multicolumn{2}{c|}{\textbf{RMSE}} \\ \cline{2-5} 
 & \textbf{Mean} & \textbf{Median} & \textbf{Mean} & \textbf{Median} \\ \hline
Baseline data (12K)    & 0.1896        & 0.1592      & 0.4876     & 0.4625 \\ \hline
Filter SSIM $\geq$ 0.1 & 0.2039     & 0.1680      & 0.4672	 & 0.4475    \\ \hline
Filter SSIM $\geq$ 0.2 & 0.2065     & 0.1730       & 0.4671	 & 0.4416 \\ \hline
Filter SSIM $\geq$ 0.3 & 0.2042     & 0.1670       & 0.4643    & 0.4400          \\ \hline
\end{tabular}
\end{table}

\begin{figure*}[h!]
    \centering
    \begin{subfigure}{0.23\textwidth}
        \includegraphics[width=\linewidth]{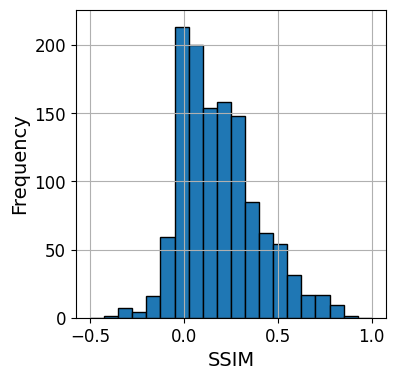}
        \caption{Baseline data}
    \end{subfigure}
    \begin{subfigure}{0.23\textwidth}
        \includegraphics[width=\linewidth]{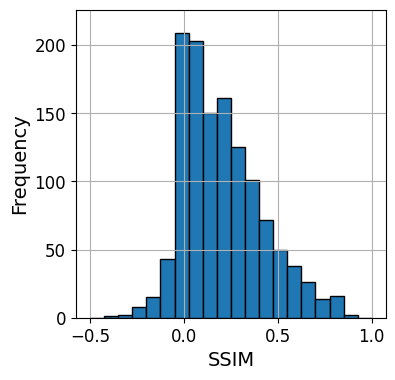}
        \caption{Filtered: SSIM $\geq$ 0.1}
    \end{subfigure}
    \begin{subfigure}{0.23\textwidth}
        \includegraphics[width=\linewidth]{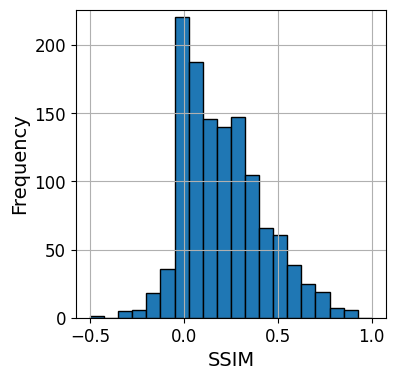}
        \caption{Filtered: SSIM $\geq$ 0.2}
    \end{subfigure}
    \begin{subfigure}{0.23\textwidth}
        \includegraphics[width=\linewidth]{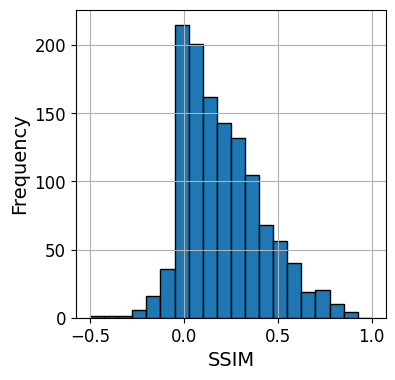}
        \caption{Filtered: SSIM $\geq$ 0.3}
    \end{subfigure}
    \caption{Histograms of SSIM distributions on the test set before and after filtering the training data at thresholds of SSIM $\geq$ 0.1, 0.2, and 0.3.}
    \label{fig:ssim_histograms}
\end{figure*}

\begin{figure}[h!]
    \centering
    \begin{subfigure}{0.22\textwidth}
        \includegraphics[width=\linewidth]{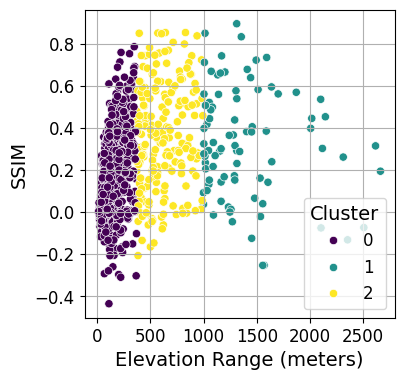}
        \caption{SSIM vs. elevation range.}
        \label{fig:ssim_range_clusters}
    \end{subfigure}
    \hfill
    \begin{subfigure}{0.22\textwidth}
        \includegraphics[width=\linewidth]{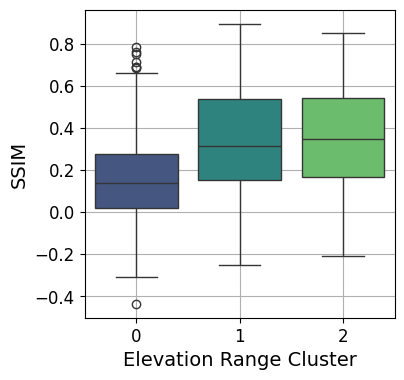}
        \caption{SSIM distribution.}
        \label{fig:ssim_boxplot_clusters}
    \end{subfigure}
    \caption{SSIM performance based on elevation range.}
    \label{fig:ssim_analysis}
\end{figure}

The SSIM histograms in \Fig~\ref{fig:ssim_histograms} provide insights into the refinement process. The histogram for the baseline dataset (\Fig~\ref{fig:ssim_histograms}a) shows a wider spread of SSIM values in the test set, with many samples clustering at lower scores. Filtering was applied only to the training data, based on SSIM thresholds of the training set, while the test set remained unchanged for consistent evaluation. After filtering the training data at thresholds of SSIM $\geq$ 0.1, 0.2, and 0.3 (\Fig~\ref{fig:ssim_histograms}b-d), the test set distributions exhibit narrower spreads, higher means, and fewer low-scoring outliers. Filtering at SSIM $\geq$ 0.2 achieves a balance between retaining high-quality samples in the training data and improving performance on the test set, validating the iterative filtering approach.

Using the refined dataset (SSIM $\geq$ 0.2), further analysis highlights the model's ability to handle various scenarios. Visual examples in \Fig~\ref{fig:sample_images} demonstrate that in mountainous regions with larger elevation ranges, the model achieves moderate SSIM values (\Fig~\ref{fig:ssim_range_clusters}). Cluster analysis shows that areas with moderate to high elevation ranges (Clusters 1 and 2) generally achieve higher SSIM scores than areas with smaller elevation ranges (Cluster 0), supporting the hypothesis that the model performs better in regions with significant elevation variations.

In contrast, regions with lower elevation ranges, often corresponding to flat or lowland areas, present challenges for the model. As shown in \Fig~\ref{fig:ssim_boxplot_clusters}, Cluster 0 has a significantly lower median SSIM compared to Clusters 1 and 2, indicating that subtle topographical variations in flatter regions are harder to capture. Despite the improved SSIM scores in Clusters 1 and 2, variability remains, likely due to complex terrain or limited training data in certain mountainous areas.


\subsection{Discussion}

\paragraph{Comparison with Previous Studies}
Compared to conventional methods of DEM generation, which often rely on LiDAR \cite{aguilar2010modelling, pinton2021estimating} or stereo-photogrammetry \cite{st2008mapping, wang2019demgeneration}, the proposed approach offers a cost-effective alternative, particularly for regions where these data sources are unavailable. The model's performance in mountainous regions aligns with findings in previous studies \cite{panagiotou2020generating, gavriil2019void, jiao2020superresolution} that generative models can effectively capture high-gradient features. However, its underperformance in lowland and residential areas underscores the need for further model refinement.

\paragraph{Implications of Findings}
The implications of this research extend to practical applications, particularly in dynamic landscapes, such as for flood modelling \cite{feliren2024progressive} and mining detection and segmentation \cite{saputra2025multi}. In most locations around the world, topography remains relatively stable, making the 2000 SRTM DEM dataset suitable for use. However, in dynamic landscapes, such as mining or urban areas, where the terrain is constantly altered by human activity, a current DEM is crucial for accurately characterizing the condition of these landscapes. Additionally, characterizing changes in the landscape over time is also important for understanding the ongoing impacts of human use. Our model can potentially generate DEMs using input data from Landsat from any point in its historic archive, starting from Landsat 5 and subsequent satellites (i.e. from 1984 onward). The outputs provide high fidelity in specific terrains, and the model can complement existing datasets and workflows. In particular, it can add valuable features for other downstream tasks such as object detection and semantic segmentation by providing elevation data as an additional band in a multi-modal deep learning manner.


\section{Conclusion}

This study demonstrates the potential of generative models, specifically the \textit{pix2pix} conditional GAN, for generating DEMs from RGB satellite imagery. The model achieved a mean RMSE of 0.4671 and a mean SSIM score of 0.2065 (scale -1 to 1), performing well in mountainous regions but facing challenges in lowland and residential areas. These findings highlight the promise of machine learning-based approaches as cost-effective alternatives to traditional DEM generation methods. Future work could focus on refining the model for lowland and urban areas by integrating additional data sources, such as multispectral or radar imagery. Expanding the application of this approach to specific use cases, such as mine detection and segmentation, could further enhance its utility for geospatial analysis.

\small
\bibliographystyle{IEEEtranN}

\bibliography{references}

\end{document}